\documentclass[%
 twocolumn,
 reprint,
showpacs,
 amsmath,amssymb,
 aps,
]{revtex4}

\usepackage{color}
\usepackage{ulem}
\usepackage{graphicx}
\usepackage{dcolumn}
\usepackage{bm}

\begin{document}

\preprint{APS/123-QED\author{Takahiro Ohgoe}}

\title{Commensurate Supersolid of Three-Dimensional Lattice Bosons}

\author{Takahiro Ohgoe$^1$}
\author{Takafumi Suzuki$^2$}
\author{Naoki Kawashima$^1$}%
\affiliation{%
	$^1$Institute for Solid State Physics, University of Tokyo, Kashiwa, Chiba 277-8581, Japan\\
	$^2$Research Center for Nano-Micro Structure Science and Engineering, Graduate School of Engineering, University of Hyogo, Himeji, Hyogo 671-2280, Japan
}%

\date{\today}

\begin{abstract}
Using an unbiased quantum Monte Carlo method, we obtain convincing evidence of the existence of a checkerboard supersolid at a {\it commensurate} filling factor 1/2 (commensurate supersolid) in the soft-core Bose-Hubbard model with nearest-neighbor repulsions on a cubic lattice. In conventional cases, supersolids are realized at incommensurate filling factors by a doped-defect-condensation mechanism, where particles (holes) doped into a perfect crystal act as interstitials (vacancies) and delocalize in the crystal order. However, in the above model, a supersolid state is stabilized even at the commensurate filling factor 1/2 {\it without doping}. By performing grand canonical simulations, we obtain a ground-state phase diagram that suggests the existence of a supersolid at a commensurate filling. To obtain direct evidence of the commensurate supersolid, we next perform simulations in canonical ensembles at a particle density $\rho=1/2$ and exclude the possibility of phase separation. From the obtained snapshots, we discuss the microscopic structure and observe that interstitial-vacancy pairs are unbound in the crystal order. 
\end{abstract}

\maketitle


	Since the observation of a nonclassical moment of inertia in solid $^{4}$He by Kim and Chan\cite{kim2004-1, kim2004-2}, the possible existence of supersolid has been discussed intensively. Regarding the possibility of a bulk supersolid in $^{4}$He, Prokof'ev and Svistunov pointed out that, in the absence of symmetry between vacancies and interstitials, a commensurate supersolid in continuous space has zero probability of being observed in nature\cite{prokofev2005}. Thus, they suggested that another scenario should be considered to interpret the decrease in nonclassical rotational inertia observed in $^{4}$He. In contrast to the case of continuous spaces, the presence of supersolid states in lattice systems has been established by unbiased quantum Monte Carlo simulations\cite{batrouni2000, wessel2005, boninsegni2005,sengupta2005, batrouni2006, suzuki2007, dang2008, yamamoto2009, pollet2010,capogrosso2010-1, xi2011}, particularly in the case when particles are added to or removed from perfect crystals. As a result of recent progress in experiments on cold atoms and molecules, optical lattice systems have become highly promising systems for realizing the supersolid state. In conventional cases, we can call such a supersolid state an incommensurate supersolid state, because it is realized by doping defects such as interstitials or vacancies into perfect crystals.

	Although a commensurate supersolid in continuous space has zero probability of being found in nature, this does not apply to systems with explicitly broken translational symmetry such as lattice systems\cite{prokofev2005}. Among the optical lattice systems, one of the most promising candidates for realizing a supersolid is a checkerboard-type supersolid near filling factor 1/2, because it can be realized only in the presence of appropriate nearest-neighbor repulsions of soft-core bosons on simple square or cubic lattices\cite{sengupta2005, yamamoto2009, xi2011}. Therefore, it is an interesting question whether such a checkerboard supersolid can be stabilized even at the commensurate filling factor 1/2 {\it in the absence of doping}. Regarding the soft-core extended Bose-Hubbard model with nearest-neighbor repulsions, the mean-field analysis\cite{yamamoto2009} and the Gutzwiller variational method\cite{otterlo1995, kimura2011} predicted that a checkerboard supersolid is stabilized at the commensurate filling factor (particle density) $\rho=1/2$ in addition to above and below $\rho=1/2$. In contrast to these studies, exact quantum Monte Carlo studies have found evidence of a supersolid phase only above $\rho=1/2$ in 1D and 2D systems\cite{batrouni2006, sengupta2005}. However, recent quantum Monte Carlo studies have revealed that a supersolid phase exists below $\rho=1/2$ as well as above $\rho=1/2$ in the case of a 3D cubic lattice\cite{yamamoto2009, xi2011}. This is consistent with the prediction by the mean-field analysis and the Gutzwiller variational method, and  also suggests the presence of a supersolid phase at $\rho=1/2$. To obtain direct evidence of a supersolid existing at the commensurate density $\rho=1/2$, it is necessary to consider the canonical ensemble at $\rho=1/2$ and exclude the possibility of a phase separation, neither of which have been investigated previously. In this Letter, using an unbiased quantum Monte Carlo method, we show convincing evidence of a commensurate supersolid existing in the above-mentioned 3D system. The finite-temperature transition to a commensurate supersolid is also investigated. Finally, we show snapshots of the commensurate supersolid and discuss a microscopic structure.
  
	The model considered here is the soft-core Bose Hubbard model with nearest neighbor repulsions on cubic lattices. The Hamiltonian is given by
	\begin{eqnarray}
		H & = & - t \sum_{\langle i, j\rangle} ( b_{i}^{\dagger} b_{j} + h.c. ) - \mu \sum_{i} n_{i} + \frac{U}{2} \sum_{i} n_{i} (n_{i}-1) \nonumber \\ & &+ V \sum_{\langle i,j \rangle} n_i n_j.
	\end{eqnarray}
	Here, $b^{\dagger}_{i}$($b_{i}$) and $n_{i}$ are the bosonic creation (annihilation) operator on site $i$ and the particle number operator defined as $n_{i} = b^{\dagger}_{i} b_{i}$, respectively. Furthermore, $t$, $\mu$, $U$, and $V$ represent the hopping parameter, the chemical potential, the on-site repulsion, and the nearest-neighbor repulsion, respectively. The summation $\langle i, j \rangle$ is taken over all pairs of nearest-neighbor sites. In our simulations, we treat $N = L^3$ systems with the periodic boundary condition. To investigate the above model, we performed unbiased quantum Monte Carlo simulations based on a hybrid algorithm of a worm algorithm\cite{prokofev1998, syljuasen2002, kato2009} and an $O(N)$ Monte Carlo method\cite{fukui2009}. This hybrid algorithm enables efficient simulation even in the presence of off-site interactions between bosons.

	\begin{figure}[t]
		\includegraphics[width=7cm]{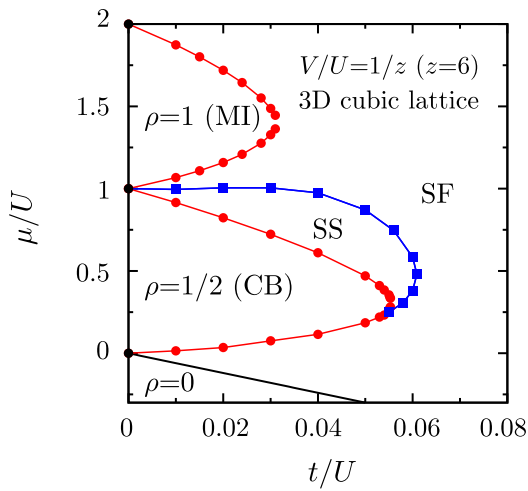}
		\caption{\label{fig:phasediag} (Color online) Ground-state phase diagram of the soft-core Bose Hubbard model with nearest neighbor repulsions on cubic lattices. Error bars are drawn but most of them are much smaller than the symbol size (here and in the following figures). Lines are used to guide the eyes. The black points at $t/U=0$ and the black solid line separating $\rho=0$ from SF are determined analytically.}
	\end{figure}

    \begin{figure}[t]
		\includegraphics[width=9.5cm]{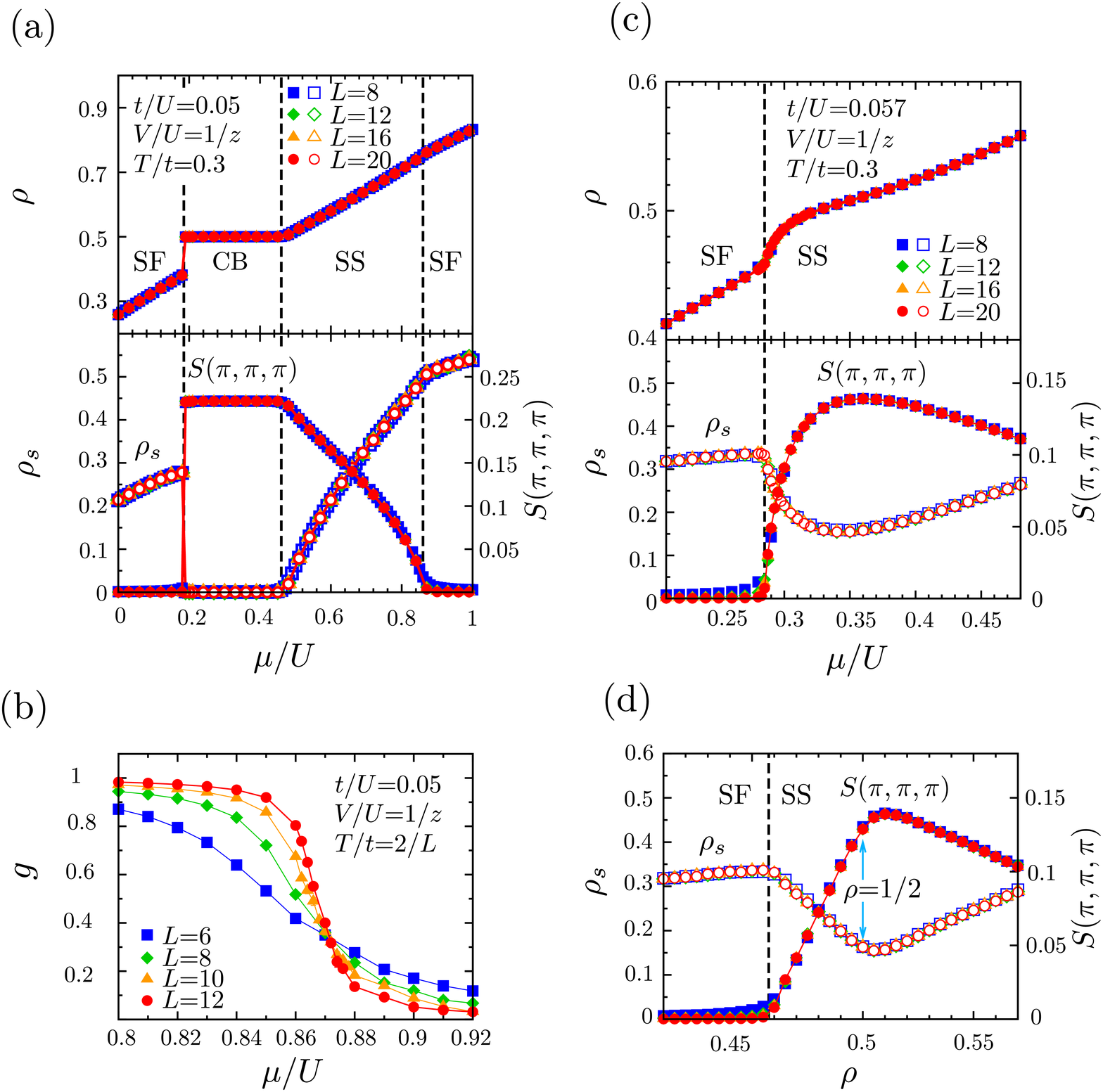}
		\caption{\label{fig:quantities} (Color online) (a) Physical quantities as functions of the chemical potential at $t/U=0.05$. (b) Determination of the quantum critical point between SS and SF from the intersection of the Binder ratios for different system sizes. (c) Physical quantities as functions of the chemical potential at $t/U=0.057$. (d) Physical quantities as functions of the particle density for the same parameter set as that in (c).}
	\end{figure}

	We show the ground-state phase diagram of the model in Fig. \ref{fig:phasediag}. In our simulations, we set the nearest-neighbor repulsion as $V/U=1/z$, where $z=6$ is the coordination number. For $0 \leq \rho \alt 1$, we confirmed the presence of a superfluid (SF) phase, a checkerboard (CB) solid phase at $\rho=1/2$, a Mott insulator (MI) phase at $\rho=1$, and a supersolid (SS) phase. The phase diagram is qualitatively in agreement with previous results obtained by the mean-field analysis\cite{yamamoto2009} and the Gutzwiller variational method\cite{otterlo1995, iskin2011}. Since the added particles on the checkerboard solid are subjected to an almost flat potential on the checkerboard-type background at $U/V \sim 1/z$, they can gain high kinetic energy and delocalize without destroying the crystal order\cite{yamamoto2009}. Consequently, a broad interstitial-based SS phase appears above the CB solid phase.

To determine the phase boundaries, we calculated the particle density $\rho = 1/N \sum_{i} \langle n_{i} \rangle$, the superfluid stiffness $\rho_{s} = \langle \mbox{\boldmath $W$}^2 \rangle T/(Lt)$, and the structure factor $S( \mbox{\boldmath $k$} ) = 1/N \sum_{i, j} e^{i \mbox{\boldmath $k$} \cdot \mbox{\boldmath $r$}_{ij}} (\langle n_{i} n_{j} \rangle - \langle n_{i} \rangle^2 )$. Here, $\mbox{\boldmath $W$}$, $\mbox{\boldmath $k$}$, and $\mbox{\boldmath $r$}_{ij}$ are the winding number vector in the path integral representation, the wave vector, and the relative position vector between sites $i$ and $j$, respectively. Furthermore, $\langle \cdots \rangle$ indicates the thermal expectation value. In Fig. \ref{fig:quantities}(a), the results are shown as functions of the chemical potential $\mu/U$ at $(t/U, V/U, T/t)=(0.05, 1/z, 0.3)$. Since the broken symmetries in CB and SF are not associated with each other, it is expected that a first-order phase transition takes place. We estimated its phase boundary from the position of the discontinuity of $\rho$, $\rho_{s}$ and $S(\pi, \pi, \pi)$. As a simple but reasonable way to determine the boundary between CB and SS, we calculated the zero-momentum Green function and estimated the energy gap for inducing the particle/hole excitation\cite{capogrosso2007}. In the same manner, the boundary of MI is also obtained from the zero-momentum Green function. The determination of the phase boundary between SS and SF requires a more qualitative consideration. As shown in Fig. \ref{fig:quantities}(b), we estimated it from the intersection of the Binder ratio $g=1/2[3-\langle m^4 \rangle/\langle m^2 \rangle^2]$ for different system sizes, where $m=1/N \sum_{i} n_{i} e^{i \mbox{\boldmath $k$} \cdot \mbox{\boldmath $r$}_{i}}$ at $\mbox{\boldmath $k$}=(\pi, \pi, \pi)$. In this analysis, we assumed that the dynamical critical exponent is equal to 1, and we fixed the temperature as $T/t \propto 1/L$. For $t/U \alt 0.03$, we observed a discontinuity in the quantities at the boundary between SS and SF. Therefore, we determined the boundary from the position of the discontinuity.

	For $t/U\agt0.055$, the SS phase covers the tip of the CB lobe. This result suggests the possible existence of commensurate supersolids, because we can expect that the system with a fixed density $\rho=1/2$ undergoes a phase transition from the CB solid state not directly to the SF state, but to the SS state as $t$ increases. This can be also expected from $\mu$ dependence of the measured quantities. In Fig. \ref{fig:quantities}(c), the physical quantities at $(t/U, V/U, T/t)=(0.057, 1/z, 0.3)$ are shown as functions of the chemical potential. As the chemical potential increases, the SF-to-SS transition occurs at $\rho<1/2$, and the particle density increases to above $\rho=1/2$. Since there is no insulating phase in which the particle density has a plateau and the superfluid stiffness vanishes, we expect that the SS phase remains even at $\rho=1/2$. To explain this in more detail, we also show $\rho_{s}$ and $S(\pi, \pi, \pi)$ as functions of the particle density $\rho$ in Fig. \ref{fig:quantities}(d). These results are obtained by making bins for particle densities with a finite width and categorizing each sample into the appropriate bin. Although these results support the hypothesis that both finite $\rho_s$ and $S(\pi, \pi, \pi)$ survive at $\rho=1/2$, we also have to exclude the possibility of a phase separation at exactly $\rho=1/2$, to conclude the presence of a supersolid at the commensurate filling factor 1/2.

	\begin{figure}[t]
		\includegraphics[width=8cm]{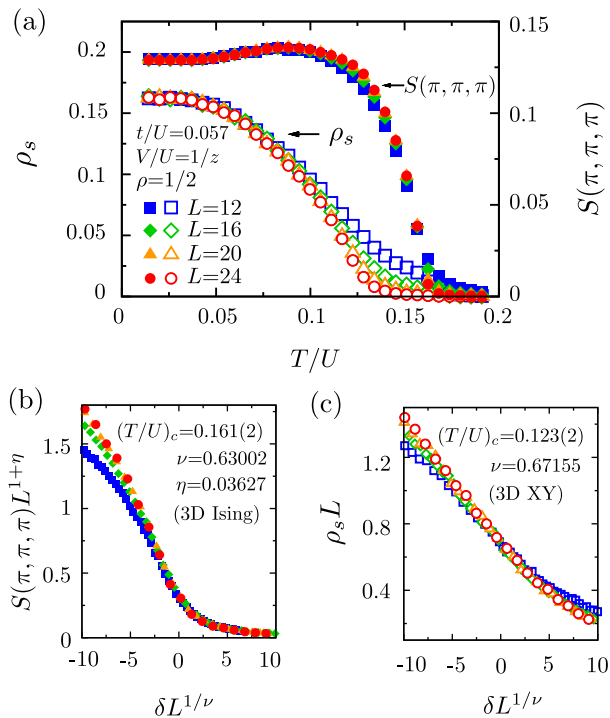}
		\caption{\label{fig:finiteT} (Color online) (a) Finite-temperature transition to the commensurate supersolid state. (b) Finite-size scaling of $S(\pi, \pi, \pi)$ for the normal-to-CB transition. (c) Finite-size scaling of $\rho_s$ for the CB-to-SS transition.}
	\end{figure}

	\begin{figure}[t]
		\includegraphics[width=7cm]{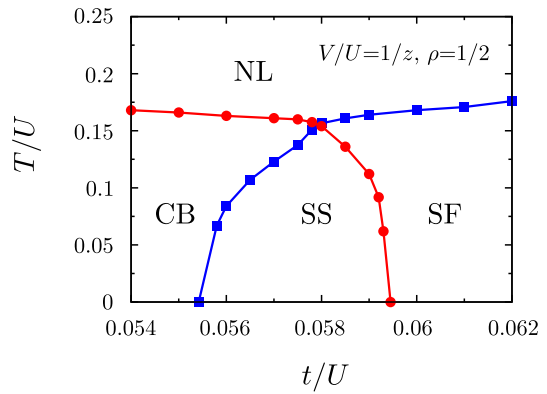}
		\caption{\label{fig:finiteTphasediag} (Color online) Finite-temperature phase diagram at the fixed density $\rho=1/2$. Circles and squares denote the critical temperatures for the checkerboard order and superfluidity, respectively. The disordered phase at finite temperatures is referred to as normal liquid (NL) phase.}
	\end{figure}

	To obtain direct evidence of the existence of a commensurate supersolid, we next perform simulations for $\rho=1/2$. The following results were obtained by using only samples with $\rho$ that is exactly equal to 1/2. In other words, we performed quantum Monte Carlo calculations in a canonical ensemble at $\rho=1/2$. In Fig. \ref{fig:finiteT}, the temperature dependences of $\rho_s$ and $S(\pi, \pi, \pi)$ are shown at $(t/U, V/U)=(0.057, 1/z)$. We can confirm finite values of $\rho_{s}$ and $S(\pi, \pi, \pi)$ at low temperatures. In our simulations, $\rho_{s}$ and $S(\pi, \pi, \pi)$ for all replicas, which were simulated under different initial conditions and with different seeds of random number, converged to the same values. This indicates the absence of a phase separation.

	To clarify the existence of a finite-temperature transition to the commensurate supersolid state, we performed finite-size scaling analysis. Since the breaking symmetries in the supersolid state are $Z_{2}$ and $U(1)$, which are related to the checkerboard order and the superfluidity, respectively, two successive transitions of the Ising-type universality class and the XY universality class are expected. We analyze the finite-size-scaling behavior of the structure factor and the superfluid stiffness by considering the scaling forms $S(\pi, \pi, \pi) L^{1+\eta}=f(\delta L^{1/\nu})$ and $\rho_{s}L=g(\delta L^{1/\nu})$, respectively. Here, $\delta$ is defined as $\delta = (T-T_{c})/T_{c}$, and $T_{c}$ is the critical temperature. Using the critical exponents of the 3D Ising universality class ($\nu$=0.63002 and $\eta$=0.03627\cite{hasenbusch2010}) for the structure factor and the 3D XY universality class ($\nu$=0.67155\cite{campostrini2001}) for the superfluid stiffness, we successfully observed data collapses. As shown in Figs. \ref{fig:finiteT}(b) and (c), the data of system sizes larger than $L\sim16$ are necessary to obtain good scaling results. These results of finite-size scaling also strongly indicate the absence of a phase separation at $\rho=1/2$. Therefore, we conclude that a true commensurate checkerboard supersolid exists in the present model. In Fig. \ref{fig:finiteTphasediag}, we summarize the transition points in the finite-temperature phase diagram at $\rho=1/2$. 
	
\begin{figure}[t]
		\includegraphics[width=8cm]{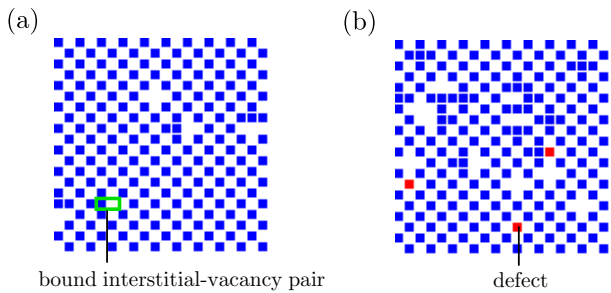}
		\caption{\label{fig:snapshot} (Color online) Cross-sectional snapshots at a fixed $z$-coordinate and imaginary time in $N$=$20^3$ systems.  (a) Checkerboard solid state at $(t/U, V/U, T/U)$=$(0.052, 1/z, 0.02)$ and (b) commensurate supersolid state at $(t/U, V/U, T/U)$=$(0.056, 1/z, 0.02)$. Particle filling is exactly $\rho=1/2$ in both cases, although this cannot be confirmed from a single cross section above. Sites are denoted as rectangles. Empty rectangles, blue rectangles, and red rectangles indicate empty sites, singly occupied sites, and doubly occupied sites, respectively. In (a), the green rectangular frame represents a bounded interstitial-vacancy (particle-hole) pair confined within a nearest-neighbor site. In (b), the existence of several doubly occupied sites suggests the presence of interstitials and vacancies that are unbound beyond the nearest-neighbor sites.}
	\end{figure}

   Finally, we discuss the microscopic structure of the commensurate supersolid. As discussed in Ref. \cite{prokofev2005}, in the supsersolid state, interstitials and vacancies do not form bound pairs and can be found arbitrarily far from each other. The delocalized defects give rise to superfluidity in crystal orders. In contrast, in the insulating phases, locally created particle-hole pairs are typically confined within nearest-neighbor sites, as has been well discussed for the Mott insulator at $\rho=1$\cite{yokoyama2011}. To qualitatively confirm the characteristic features, we show snapshots at the commensurate density $\rho=1/2$ in Fig. \ref{fig:snapshot}. In the checkerboard insulating phase [Fig. \ref{fig:snapshot}(a)], we observe bound interstitial-vacancy pairs that are typically confined within nearest-neighbor sites. In contrast, in the commensurate supersolid state [Fig. \ref{fig:snapshot}(b)], we can easily observe interstitials and vacancies that are separated by a distance exceeding the nearest-neighbor distance. This can be confirmed from the presence of doubly occupied sites because such sites are never created from interstitial-vacancy pairs confined within nearest-neighbor sites. Since we did {\it not} dope any interstitials or vacancies into the commensurate solid, the defects originated from the unbound interstitial-vacancy pairs. In this sense, it is expected that the mechanism of the CB-to-SS transition is similar to the released doublon-holon mechanism of the MI-to-SF transition at $\rho=1$\cite{yokoyama2011}.

	In conclusion, using exact quantum Monte Carlo simulations, we obtained direct evidence of the existence fo a checkerboard supersolid at $\rho=1/2$ in 3D bosonic systems. By obtaining a finite-temperature phase diagram at $\rho=1/2$, we also showed the range of parameters in which this supersolid exists, including its temperature. Finally, from snapshots, we qualitatively discussed the its microscopic structure. Although the snapshots suggest the existence of unbound interstitial-vacancy pairs in the commensurate supersolid, further investigation is needed to confirm this quantitatively. 

	This work was financially supported by the Global COE Program ``the Physical Science Frontier", a Grant-in-Aid for Scientific Research (B) (22340111), and a Grant-in-Aid for Scientific Research on Priority Areas ``Novel States of Matter Induced by Frustration" (19052004) from MEXT, Japan. The simulations were performed on computers at the Supercomputer Center, Institute for Solid State Physics, University of Tokyo.


\end{document}